\documentclass[11pt,preprint]{aastex}

\usepackage{natbib}
\newcommand{\kms}{~km~s$^{-1}$} 
\newcommand{\teff}{$T_{\rm eff}$}
\newcommand{\logg}{$\log g$}

\newcommand{\vt}{$v_{\rm micro}$}

\slugcomment{Li in double-lined spectroscopic binary}

\shorttitle{Li abundances of G~166--45} 
\shortauthors{Aoki et al.}

\begin{document}

\title{Examination of the mass-dependent Li depletion hypothesis by
  the Li abundances of the very metal-poor double-lined spectroscopic 
  binary G~166--45}



\author{Wako Aoki \altaffilmark{1,2}}
\affil{National Astronomical Observatory of Japan, Mitaka, Tokyo,
181-8588 Japan}
\email{aoki.wako@nao.ac.jp}
\affil{Department of Astronomical Science, The Graduate
  University of Advanced Studies, Mitaka, Tokyo, 181-8588 Japan}
\author{Hiroko Ito \altaffilmark{2}}
\author{Akito Tajitsu \altaffilmark{3}}
\affil{The Subaru Telescope, National Astronomical Observatory of Japan, 650 North A'ohoku Pl., Hilo, HI 96720}





\begin{abstract}

The Li abundances of the two components of the very metal-poor
([Fe/H]$=-2.5$) double-lined spectroscopic binary G~166--45
(BD+26$^{\circ}$2606) are determined separately based on high
resolution spectra obtained with the Subaru Telescope High Dispersion
Spectrograph and its image slicer. From the photometric colors and the
mass ratio the effective temperatures of the primary and secondary
components are estimated to be 6350$\pm 100$~K and 5830$\pm 170$~K,
respectively. The Li abundance of the primary ($A$(Li)$=2.23$) agrees well 
with the Spite plateau value, while that of the secondary is slightly
lower ($A$(Li)$=2.11$). Such a discrepancy of the Li
abundances between the two components is previously found in the
extremely metal-poor, double-lined spectroscopic binary CS~22876--032,
however, the discrepancy in G~166--45 is much smaller. The results
agree with the trends found for Li abundance as a function of
effective temperature (and of stellar mass) of main-sequence stars
with $-3.0<$[Fe/H]$<-2.0$, suggesting that the depletion of Li at
{\teff}$\sim 5800$~K is not particularly large in this metallicity range.
The significant Li depletion found in CS~22876--032B is a
phenomenon only found in the lowest metallicity range ([Fe/H]$<-3$).


\end{abstract}





\section{Introduction}

Lithium is an important element as a diagnostic of the structure and
evolution of low-mass stars. Since Li is easily destroyed by nuclear
fusion in the hot interiors of stars where the temperature is above
2.5 $\times 10^{6}$~K, stellar internal processes can be examined from
observed Li abundances at stellar surface. Li depletion processes also
play an important role in reconciling the serious discrepancy between
the observed constant Li abundance, the so-called Spite plateau, of
metal-poor turnoff stars ($A$(Li) = $\log$(Li/H)+12 $\sim$ 2.2: e.g.,
\citet{spite82, asplund06}) and the primordial Li abundance predicted
from standard Big Bang nucleosynthesis (SBBN) models adopting the
baryon density determined from observations of the cosmic microwave
background (CMB) by the WMAP satellite ($A$(Li) $= 2.72$: e.g. Cyburt
et al. 2008). The discrepancy seems to be larger at lower metallicity
\citep{bonifacio07, aoki09, sbordone10}.  If the stellar intrinsic Li
depletion is in operation in the atmospheres of metal-poor turnoff
stars, the observed Li abundance is lower than the initial value, which
may have been closer to the CMB and SBBN prediction.

Although no Li depletion is inferred for metal-poor stars on the
main-sequence-turn-off by standard stellar evolution theories
because of their shallow convection zones, evolutionary models that
include effects of atomic diffusion \citep[e.g.,
][]{richard02a, richard02b} predict significant decline of
surface Li abundances for metal-poor dwarfs. \citet{richard05}
additionally introduced parametrized effects of turbulent mixing to
their evolution model for [Fe/H] $= -2.31$ \footnote{[A/B] = $\log(N_{\rm
A}/N_{\rm B}) -\log(N_{\rm A}/N_{\rm B})_{\odot}$, and $\log\epsilon_{\rm A}
=\log(N_{\rm A}/N_{\rm H})+12$ for elements A and B.}, though the understanding
of the physics of the turbulent mixing is still
insufficient. \citet{korn06} and \citet{korn07} investigated effects
of these processes for the globular cluster NGC~6397 by examining
small variations of Mg, Ca, Ti and Fe abundances in main-sequence
turnoff, subgiant and giant stars. They suggested diffusion process
during main-sequence phase and attributed the slightly lower Li
abundances in turn-off stars than in subgiants to these effects.

The most convincing evidence of stellar Li depletion in metal-poor
dwarfs found so far is the large difference between Li abundances of
the primary and secondary dwarf stars in the extremely metal-poor
([Fe/H]$ = -3.6$) double-lined spectroscopic binary CS~22876--032,
which is about 0.4 dex \citep{gonzalezhernandez08}. Since the two
stars should have shared the same chemical compositions at their
birth, it is reasonable to presume that depletion processes have made
the Li abundance difference during the stellar evolution.
Such depletion is not expected for a star with the effective
temperature of the secondary ({\teff}$=5900$~K) from previous studies
of less metal-poor stars. Given that the mass of the secondary star
is about 0.9 times that of the primary, Li depletion may operate
mass-dependently, which is consistent with the predictions of
\citet{richard05}, as discussed by \citet{melendez10}.

The best approach to examine the hypothesis of the mass-dependent
lithium depletion is accurate determinations of Li abundances for
several metal-poor binary stars. Although \citet{melendez10}
suggested a correlation between Li abundance and stellar mass for
metal-poor single stars, investigation of single stars which have
different ages and distances would suffer from uncertainties in mass
estimates.

In this Letter, we report our chemical abundance analysis, in
particular the Li abundances, of the two components of the very
metal-poor, double-lined spectroscopic binary G~166--45. This object
is selected from the sample of \citet{goldberg02} who investigated the
orbital parameters of 34 binary stars found by the Carney-Latham
proper motion sample \citep{carney94}. The chemical abundances of
individual components of this object have not yet been
investigated. We discuss in particular the implication of the small
difference of the Li abundance between the two components found from
the present analysis of high resolution spectra.

\section{Observations and measurements}


A high-resolution spectrum of G~166--45 was obtained with the Subaru
Telescope High Dispersion Spectrograph \citep[HDS;][]{noguchi02} on
July 22, 2011. We applied the image slicer recently installed in the
spectrograph, which provides the resolving power of $R=110,000$ with
high efficiency \citep{tajitsu12}. The wavelength range from 5100 to
7800~{\AA} is covered by the standard setup of StdRa. The exposure
time is 60 minutes, resulting in the S/N ratio (per 0.9~{\kms} pixel)
of 350 at 6700~{\AA}.

According to the orbital parameters of the binary determined by
\citet{goldberg02}, the velocity difference between the two components
of this object is expected to be 15--20~{\kms} at this epoch. Indeed,
the separation of the absorption features in our spectrum is
18.5~{\kms} (see below), agreeing well with the above estimate.

Standard data reduction was carried out with the IRAF
echelle package\footnote{IRAF is distributed by the National Optical
  Astronomy Observatories, which is operated by the Association of
  Universities for Research in Astronomy, Inc. under cooperative
  agreement with the National Science Foundation.}. Spectra are
extracted from individual sliced images and combined for the slice
numbers 2--5 in the definition of \citet{tajitsu12} to obtain the best
quality spectrum.


Equivalent widths ($W$'s) for isolated absorption lines were measured
by fitting Gaussian profiles for each component of the binary. The
equivalent widths are used in the
following chemical abundance analysis by dividing the fraction of the
contribution of each component to the continuum flux estimated from
colors and mass ratio of the two components (\S~\ref{sec:ana}).


\section{Analysis and results}\label{sec:ana}

\subsection{Atmospheric Parameters}

In order to determine the atmospheric parameters of the two components
of the binary system and their contributions to the continuum flux, we
adopt the mass ratio of the two components ($q=m_{2}/m_{1}=0.89\pm
0.04$), where $m_{1}$ and $m_{2}$ are the primary and secondary
masses, respectively, determined by \citet{goldberg02}, and calculate
the colors ($B-V$, $V-R$, and $V-I$) as a function of the primary mass
using the $Y^{2}$ isochrone \citep{demarque04} for the age of
12~Gyr. The metallicity is assumed to be [Fe/H]$=-2.5$, which is
consistent with the results obtained from the following analysis.

\begin{figure}
\epsscale{.80}
\plotone{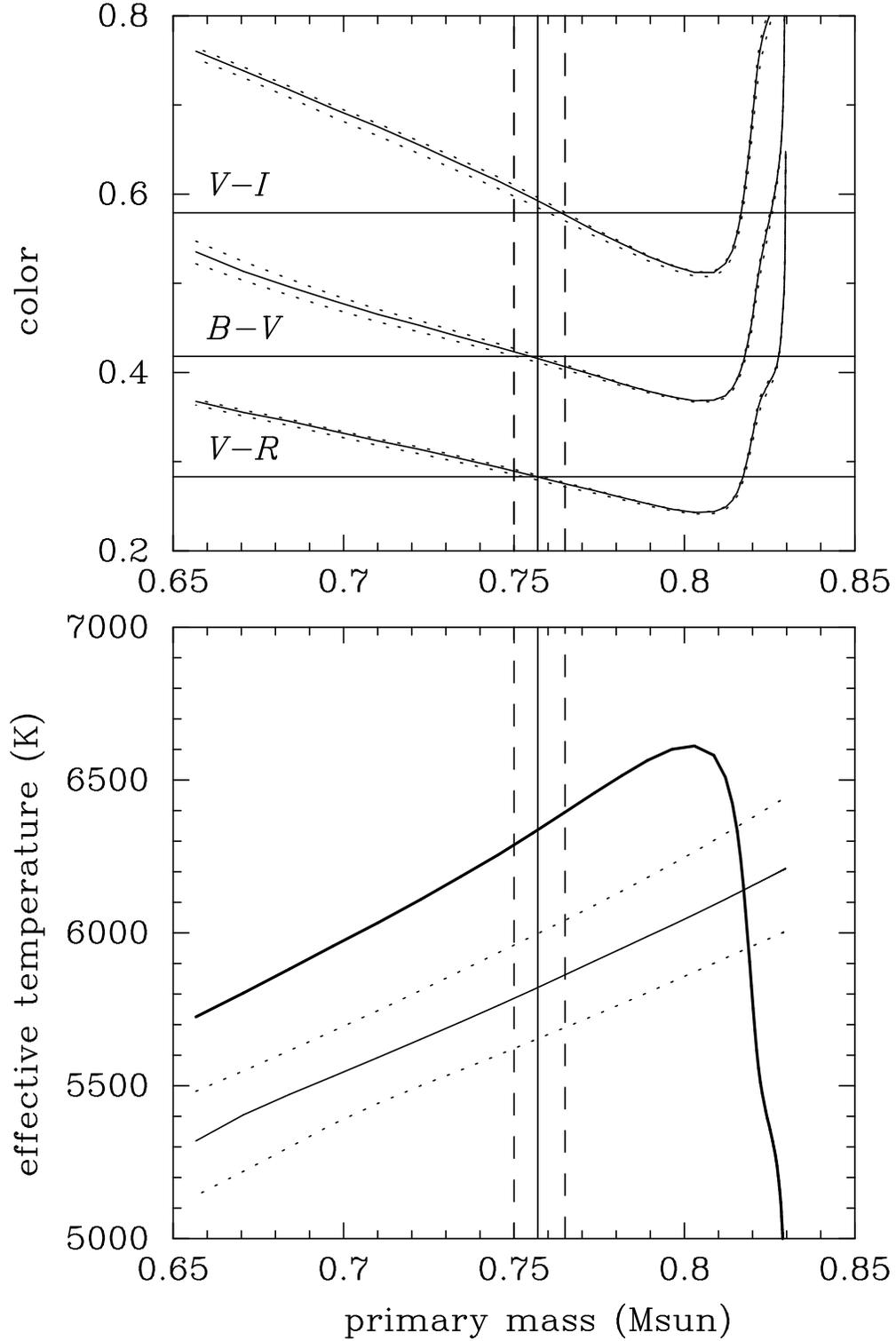}
\caption{Upper panel: Colors ($V-I, B-V, V-R$) calculated from the
  isochrones assuming mass ratios of $q=0.89\pm 0.04$. The reddening
  corrected colors of G~166--45 are shown by horizontal lines. The
  vertical solid and dashed lines indicate the adopted primary mass
  and its possible range estimated from the three colors. Lower panel:
  effective temperatures of primary (thick solid line) and secondary
  (thin solid line and dotted lines) as a function of assumed primary
  mass adopting the above mass ratios. \label{fig:param}}
\end{figure}

The upper panel of Figure~\ref{fig:param} shows the three colors
calculated using the isochrone. Solid and dotted lines indicate the
cases of $q=0.89$ and those with 0.04 higher or lower,
respectively. Very accurate colors of G~166--45 are available, because
this object is a Landolt's photometric standard star. The colors
measured by their recent work \citep{landolt07} with a reddening
correction are shown by the horizontal lines in the figure. The
reddening correction ($E(B-V)=0.02$) is estimated from the
interstellar \ion{Na}{1} D$_{2}$ line detected in our spectrum (the
equivalent width of 80~m{\AA}).

Comparisons of these colors with the calculations provide constraints
on the primary mass of this binary system (as well as the secondary
mass). The primary mass estimated from the three colors is $0.757\pm
0.007$M$_{\odot}$, in which the error corresponds to the differences
of the results from the three colors. Even if the photometry error
(0.015~mag) for G~166--45 estimated by \citet{landolt07} (see the
erratum of the paper) and the uncertainty of the mass ratio are
included, the error size of the primary mass does not change
significantly. The other solution that satisfies the colors and the
mass ratio is the case that the primary is a subgiant ($m_{1}\sim
0.82~$M$_{\odot}$) and the secondary is a main-sequence star with similar
temperatures (see Fig.~\ref{fig:param}). This solution is, however,
excluded because the luminosity difference of the two components can
not be so large, given the strengths of the absorption features of the
two components.

The lower panel of Figure~\ref{fig:param} shows the effective
temperatures of the primary and secondary, as a function of the
primary mass (assuming the mass ratio), adopted from the $Y^{2}$
isochrone. The effective temperature of the primary is well determined
to be 6350$\pm 100$~K. The error reflects the uncertainty of the mass
determination from different color indices and photometric errors, but
does not include the uncertainty of the temperature scale (depending
on model atmospheres).

By contrast, the uncertainty of the effective temperature of the
secondary is relatively large (170~K) due to the uncertainty of the
mass ratio. We adopt 5830~K for the secondary as the best estimate in
the following analysis, but also conduct the analyses for {\teff} =
5650 and 6000~K to estimate the errors due to the uncertainty of the
mass ratio.  The surface gravities of the primary and the secondary
are estimated from the isochrone to be $\log g=4.4$ and 4.6,
respectively.

The fraction of the contributions of each component to the continuum
flux in the $R$-band ($f$: $f_{\rm A}$ and $f_{\rm B}$ for the primary
and secondary, respectively), in which the \ion{Li}{1} resonance line
exists, is also estimated from the isochrone for the
three choices of the secondary's effective temperature.

The adopted parameter sets ({\teff}, {\logg}, and $f_{\rm B}$) are
(5830~K, 4.6, 0.27), (5650~K, 4.6, 0.21) and (6000~K, 4.6, 0.34)
for the secondary, while {\teff} = 6350~K, {\logg} = 4.4, and
$f_{\rm A}=1-f_{\rm B}$ are adopted for the primary (Table~1).

For comparison purposes, we applied the above procedure to the
double-lined spectroscopic binary CS~22876--032 adopting the mass
ratio of 0.91 \citep{gonzalezhernandez08} and colors given by
\citet{norris00}. The results agree well with those determined by
\citet{gonzalezhernandez08}, i.e., 6500~K and 5900~K for the primary
and secondary, respectively, with uncertainties of about 100~K.

\subsection{Chemical abundances}

Chemical compositions of the two components of this binary system are
determined separately based on the standard LTE analysis of the
equivalent widths obtained by dividing the measured
equivalent widths from the spectrum by the $f$. The ATLAS NEWODF model
atmosphere grid \citep{castelli03} assuming $\alpha$ elements excesses
is used in the analysis.

The micro-turbulent velocity ({\vt}) is determined by demanding no
dependence of the abundance results for individual \ion{Fe}{1} lines on their
strengths (equivalent widths). The derived {\vt} is 1.5~{\kms} for the
primary and 1.0~{\kms} for the secondary. 

The results of the abundance analyses are given in
Table~\ref{tab:abundance} for eight elements (along with the Li
abundance determined separately) for the standard case of the stellar
parameters. We also give the results of Fe abundances for the other
parameter sets in the table. The errors given in
Table~\ref{tab:abundance} include random errors and those due to
uncertainties of atmospheric parameters. We estimate the random errors in the
measurements to be $\sigma/N$, where $\sigma$
is the standard deviation of derived abundances from individual lines
and $N$ is the number of lines used. When only a few
lines are available, the $\sigma$ of \ion{Fe}{1} is adopted in the
estimates. The errors due to the uncertainty of the atmospheric
parameters ($\delta${\teff}$= 100$~K, $\delta${\logg}$=0.3$, and
$\delta${\vt}$= 0.2$~{\kms}) are also estimated and added in
quadrature to the random errors. The errors for the secondary could be
slightly larger if the the uncertainty of effective temperature due to
the uncertainty of the mass ratio ($\sim 170$~K) is included.

The Fe abundances derived from individual \ion{Fe}{1} lines show no
statistically significant dependence on their lower excitation
potentials for the three choices of the stellar parameters. The Fe
abundance derived from \ion{Fe}{2} lines is slightly higher than
that from \ion{Fe}{1} lines, suggesting the {\logg} values adopted in
the analysis are systematically too high (by 0.1--0.3~dex), or
the \ion{Fe}{1} lines suffer from non-LTE effects \citep[e.g., ][]{asplund05}. 
We note that the Li line analysis reported in the following subsection  is not
sensitive to the changes of the adopted {\logg} and {\vt}.

The chemical compositions of the two components agree very well with each
other. The changes of the Fe abundance from \ion{Fe}{1} lines by
changing the choice of stellar parameter set are only 0.05~dex for the
primary. The change for the secondary is even smaller, because the
effect of the change of {\teff} on the derived Fe abundance is
partially compensated by the change of the contribution of the
secondary to the continuum flux (adopting a higher {\teff} results in
a higher Fe abundance in general, while a larger $f_{\rm B}$ results
in smaller equivalent widths and a lower abundance).

\subsection{Li abundances}

\begin{figure}
\epsscale{.80}
\plotone{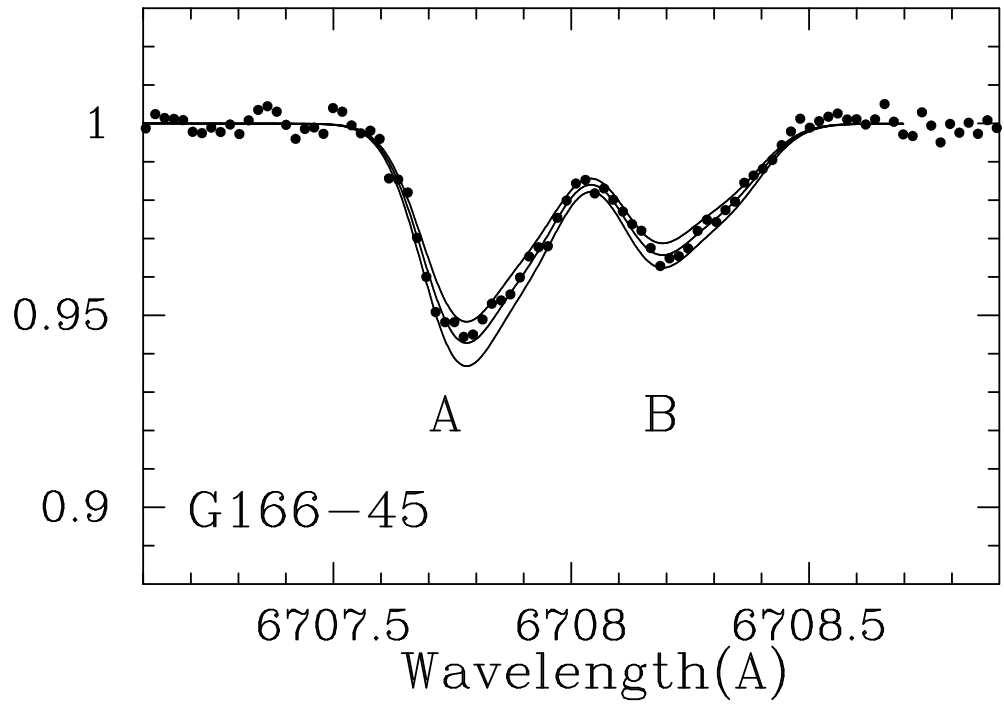} 
\caption{Comparisons of synthetic spectra for the Li line with the observed one for the two components of G~166--45. Three different Li abundances ($\pm
0.05$~dex in $A$(Li)) for the results for the standard parameter set (see text)
are assumed for each component in the calculations.
\label{fig:li}}
\end{figure}

The Li abundances of the two components of G~166--45 are determined by
fitting synthetic spectra to the observed one. The synthetic spectra
are calculated using the same model atmospheres used in the above
subsection for both components, and are combined taking their $f$'s into
account and the velocity shift between them (18.5~{\kms}). Comparisons
of the synthetic spectra with the observed one are depicted in
Figure~\ref{fig:li}, where the doppler correction is made for the
primary. In the fitting process, the $\chi^{2}$ minimum is searched
for each component: the wavelength ranges 6707.6--6708.0~{\AA} and
6708.0--6708.4~{\AA} are used to determine the Li abundances of the
primary and secondary, respectively.  The $2 \sigma$ error in the
fitting is 0.02 and 0.03~dex for the primary and secondary, respectively.
 
The results for the three choices of mass ratios are given in
Table~\ref{tab:abundance}. Interestingly, the Li abundance of the secondary is
insensitive to the choice of the mass ratio for the same reason for the behavior of the Fe abundance mentioned above.


The derived Li abundance of the secondary is 0.05--0.15~dex lower than
that of the primary. In the following discussion on the Li abundance
difference between the two components, we take the fitting errors
(0.02--0.03~dex) and the errors due to the uncertainty of the mass
ratio (0.02--0.05~dex) into consideration. We note that this does not
include the error due to the uncertainty of {\teff} scales, which is as
large as 0.1~dex in $A$(Li), but is a systematic error affecting the
results for both components.


\section{Discussion and concluding remarks}

\begin{figure}
\epsscale{.70}
\plotone{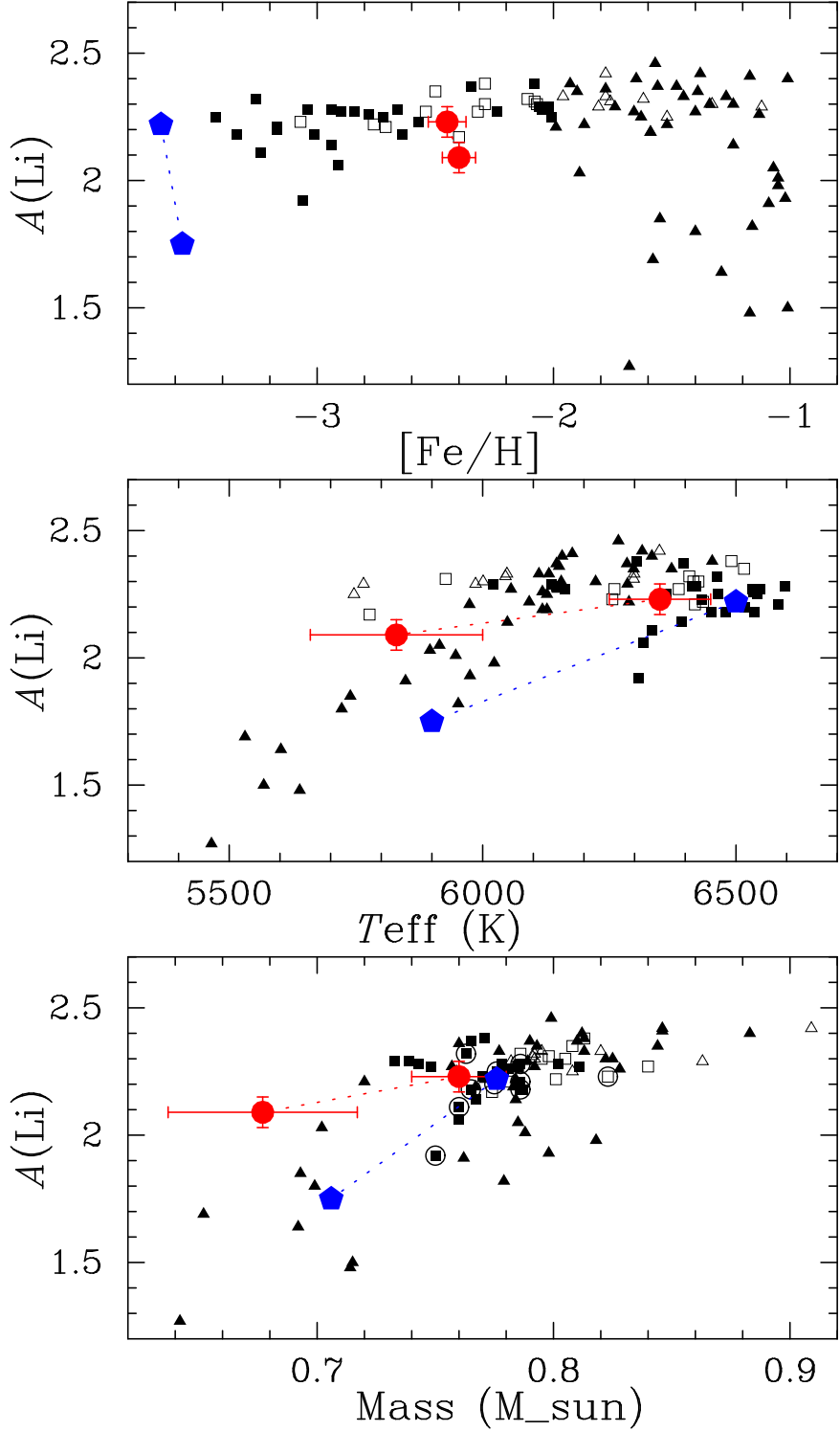}
\caption{Li abundances as functions of [Fe/H], {\teff} and stellar
  mass. Filled circles with error bars indicate G~166--45~A and B,
  while filled pentagons mean CS~22876--032A and B
  \citep{gonzalezhernandez08}: the secondary stars have lower Li
  abundances. Filled squares and triangles are 'main-sequence' ($\log
  g \geq 4.0$) stars with [Fe/H]$<-2$ and $>-2$, respectively, and
  open squares and triangles are 'subgiant' ($\log g<4.0$) stars with
  [Fe/H]$<-2$ and $>-2$, respectively \citep{melendez10}. In the
  bottom panel, objects with [Fe/H]$<-3$ are indicated by over-plotting
  large open circles. \label{fig:disc}}
\end{figure}

The Li abundance of the primary derived by the above analysis
($A$(Li)$=2.23\pm0.05$) well agrees with the Spite plateau value,
while that of the secondary is slightly lower. Figure~\ref{fig:disc}
(top panel) depicts the Li abundances of main-sequence ({\logg}$>4.0$:
filled symbols) and subgiant ({\logg}$<4.0$: open symbols) stars as a
function of [Fe/H]. The data other than G~166--45 are adopted from
\citet{gonzalezhernandez08} for CS~22876--032 and \citet{melendez10}
for other stars determined by LTE analyses. The possible Li depletion
in the main-sequence phase can be inspected from filled symbols shown
here. While large scatter of Li abundances is found in [Fe/H]$>-2$,
which is due to the lower abundances in stars with lower {\teff} (see
below), there is no distinctive scatter in $-3<$[Fe/H]$<-2$ in this
sample. This is not changed even if subgiants ({\logg}$<4$) are
included. \citet{melendez10} suggested the existence of two plateaus
with slightly higher and lower Li abundances in [Fe/H]$>-2.5$ and
$<-2.5$, respectively. The Li abundance of G~166--45A is apparently in
agreement with the lower plateau. This result is, however, not
definitive, given a possible systematic offset in Li abundance
determinations (as large as 0.1~dex) mostly due to {\teff} scales
used. Among the stars in this metallicity range, the Li abundance of
G~166--45B is relatively low.

Some scatter of Li abundances is also found in [Fe/H]$<-3$, though
that is smaller than that in [Fe/H]$>-2$. The discrepancy of the two
components of CS~22876--032 is consistent with the size of the scatter.

The middle panel of Figure~\ref{fig:disc} shows the Li abundances as a
function of {\teff}. A clear dependence of Li abundances on {\teff} is
found in {\teff}$<6000$~K. It should be noted that all stars in this
temperature range, except for CS~22876--032B and two subgiants, are
objects with [Fe/H]$>-2$. The difference of Li abundances between
objects with {\teff}$\sim 6350$~K and those with $\sim$5830~K is as
large as 0.4~dex. The discrepancy of the Li abundances found between
the two components of G~166--45 is much smaller than this value. This
suggests that the depletion of Li in stars with {\teff}$\sim 5800$~K
is much smaller at this metallicity ([Fe/H]$\sim -2.5$) than that in
[Fe/H]$>-2$. This might be related to the shallower convection zone in
stars with lower metallicity, however, other causes for Li depletion
would be required as mentioned by \citet{richard05}, who demonstrated
that the temperature of the bottom of the convective zone is almost
independent of metallicity. It should also be noted that subgiants
({\logg}$<4.0$) with {\teff}$<6000K$ have Li abundances as high as
stars with {\teff}$>6000$~K, hence, the correlation between the Li
abundance and {\teff} discussed above is only found in main-sequence
stars with high [Fe/H] ($>-2$). The exception is CS~22876--032B, which
shows a Li abundance along with the correlation found for stars with
[Fe/H]$>-2$. The reason for the Li depletion in this object is,
however, likely different from the {\teff}-dependent depletion found
for [Fe/H]$>-2$ (see below).

The bottom panel of Figure~\ref{fig:disc} shows Li abundances as a
function of stellar mass. Li depletion in lower mass stars is
discussed by \citet{melendez10}. Excluding the stars with [Fe/H]$>-2$
(indicated by triangles), which show a tight correlation between Li
abundances and {\teff}, the dependence of the Li abundance on stellar
mass is not very clear. Indeed, \citet{melendez10} reported that the
slopes of Li abundances as a function of stellar masses is significant
at only 1$\sigma$ level in [Fe/H]$<-2.5$, which increases to the
3$\sigma$ level by including the binary CS~22876--032. However, if
stars with lowest metallicity ([Fe/H]$<-3$) are selected, a
correlation seems to appear again. In order to demonstrate this, stars
in this metallicity range are indicated by over-plotting open circles
(except for CS~22876--032) in the bottom panel of the figure.

On the other hand, no clear correlation is found between the Li
abundances and stellar masses for objects with $-3<$[Fe/H]$<-2$
(squares without open circles). Our study of the Li abundance of
G~166--45~B extends this trend to a mass below 0.7~M$_{\odot}$.

 
The dependence of Li abundances on stellar mass and metallicity is
nonlinear. In [Fe/H]$>-2$, Li abundances are lower in main-sequence
stars with lower {\teff} (lower mass) in {\teff}$\lesssim 6000$~K
($M\lesssim 0.8$M$_{\odot}$). A similar dependence is also found in
[Fe/H]$<-3$, possibly in stars with even higher {\teff} ($\lesssim
6300$~K, which corresponds to $M\lesssim 0.8$M$_{\odot}$). On the
other hand, such a dependence is not clearly seen in
$-3<$[Fe/H]$<-2$. The double-lined spectroscopic binaries
CS~22876--032 and G~166--45 most clearly represent these trends in
[Fe/H]$<-3$ and $-3<$[Fe/H]$<-2$, respectively. Since the initial
abundances of the two components of a binary system can be assumed to
be the same, the discrepancy of the Li abundances found between the
two components indicates a depletion of Li in the secondary. This
suggests that the scatter of Li found in the lowest metallicity range
is due to depletion of Li in some object, which is not clearly seen in
$-3<$[Fe/H]$<-2$.


\acknowledgments




{\it Facilities:} \facility{Subaru(HDS)}.

\acknowledgments

We are grateful to M. Parthasarathy for his useful comments for the
manuscript. W.A. was supported by the Grants-in-Aid for Science
Research of JSPS (20244035).

\clearpage






\begin{deluxetable}{lcccccccc}
\tablecaption{STELLAR PARAMETERS AND CHEMICAL ABUNDANCES \label{tab:abundance}}
\tablewidth{0pt}
\tablehead{
        & & \colhead{G~166--45A} &  &  & &\colhead{G~166--45B} &&  }
\startdata
$q=0.89$ & \multicolumn{4}{l}{{\teff}=6350~K, $f_{\rm A}=0.73$}    &  \multicolumn{4}{l}{{\teff}=5830~K, $f_{\rm B}=0.27$} \\
\hline
Species & $\log\epsilon$ & [X/Fe]([Fe/H]) & $n$ & error & $\log\epsilon$ & [X/Fe]([Fe/H]) & $n$ & error \\
\hline
Fe I  & 5.05 & $-2.45$ & 35 & 0.10 & 5.10 & $-2.40$ & 39 & 0.11 \\
Fe II & 5.15 & $-2.35$ &  3 & 0.14 & 5.30 & $-2.20$ &  2 & 0.15 \\
Na I  & 3.74 & $-0.05$ &  2 & 0.09 & 3.76 & $-0.08$ &  2 & 0.09 \\
Mg I  & 5.44 & $0.29 $ &  3 & 0.09 & 5.31 & $0.11 $ &  3 & 0.12 \\
Ca I  & 4.22 & $0.34 $ &  8 & 0.08 & 4.20 & $0.26 $ &  8 & 0.08 \\
Ti II & 3.00 & $0.50 $ &  3 & 0.13 & 3.01 & $0.46 $ &  3 & 0.13 \\
Cr I  & 3.13 & $-0.06$ &  3 & 0.10 & 3.14 & $-0.10$ &  3 & 0.10 \\
Ni I  & 3.90 & $0.13 $ &  1 & 0.13 & 4.17 & $0.35 $ &  2 & 0.13 \\
Ba II &$-$0.47&$-0.20$ &  1 & 0.14 & \nodata & \nodata &  \nodata & \nodata \\
Li I  & 2.23 & \nodata &  1 & 0.02\tablenotemark{a} & 2.11 & \nodata &  1 & 0.03\tablenotemark{a} \\
\hline
$q=0.93$ & \multicolumn{4}{l}{{\teff}=6350~K, $f_{\rm A}=0.66$} &  \multicolumn{4}{l}{{\teff}=6000~K, $f_{\rm B}=0.34$} \\
\hline
Fe I  & 5.11 & $-2.39$ & 35 & 0.10 & 5.09 & $-2.42$ & 39 & 0.11 \\
Fe II & 5.13 & $-2.37$ &  3 & 0.14 & 5.19 & $-2.31$ &  2 & 0.15 \\
Li I & 2.27 & \nodata  &  1 & 0.02\tablenotemark{a} & 2.12 & \nodata &  1 & 0.03\tablenotemark{a} \\
\hline
$q=0.85$ & \multicolumn{4}{l}{{\teff}=6350~K, $f_{\rm A}=0.79$} &  \multicolumn{4}{l}{{\teff}=5670~K, $f_{\rm B}=0.21$} \\
\hline
Fe I  & 5.00 & $-2.50$ & 35 & 0.10 & 5.14 & $-2.36$ & 39 & 0.11 \\
Fe II & 5.04 & $-2.46$ &  3 & 0.14 & 5.43 & $-2.07$ &  2 & 0.15 \\
Li I & 2.18  & \nodata &  1 & 0.02\tablenotemark{a} & 2.13 & \nodata &  1 & 0.03\tablenotemark{a} \\
\enddata
\tablenotetext{a}{The errors of Li abundances represent random errors in the fitting of synthetic spectra.} 
\end{deluxetable}

\end{document}